\documentclass[fleqn,usenatbib]{mnras}

\usepackage{newtxtext,newtxmath}

\usepackage[T1]{fontenc}

\DeclareRobustCommand{\VAN}[3]{#2}
\let\VANthebibliography\thebibliography
\def\thebibliography{\DeclareRobustCommand{\VAN}[3]{##3}\VANthebibliography}

\usepackage{graphicx}	
\usepackage{amsmath}	
\usepackage{bm}

\usepackage{orcidlink}

\title[Superorbital variations in Be/X-ray binaries]{Disc precession in Be/X-ray binaries drives superorbital variations of outbursts and colour  }

\author[R. G. Martin \& P. A. Charles ]{
Rebecca G. Martin\thanks{E-mail: rebecca.martin@unlv.edu}$^{1,2}$\orcidlink{0000-0003-2401-7168}
and Philip A. Charles$^{3,4}$
\\
$^1$Nevada Center for Astrophysics, University of Nevada, Las Vegas,
4505 South Maryland Parkway, Las Vegas, NV 89154, USA\\
$^2$Department of Physics and Astronomy, University of Nevada, Las Vegas,
4505 South Maryland Parkway, Las Vegas, NV 89154, USA\\
$^3$Department of Physics \& Astronomy, University of Southampton, Southampton SO17 1BJ, UK\\
$^4$Astrophysics, Department of Physics, University of Oxford, Keble Road,  Oxford OX1 3RH, UK
}

\date{Accepted XXX. Received YYY; in original form ZZZ}

\pubyear{2023}

\begin{document}
\label{firstpage}
\pagerange{\pageref{firstpage}--\pageref{lastpage}}
\maketitle

\begin{abstract}
Superorbital periods that are observed in the brightness of Be/X-ray binaries may be driven by a misaligned and precessing Be star disc. We examine how the precessing disc model  explains the superorbital variation of (i) the magnitude of the observed X-ray outbursts and (ii) the observed  colour. 
With hydrodynamical simulations we show that the magnitude of the average accretion rate on to the neutron star, and therefore the X-ray outbursts, can vary by over an order of magnitude over the superorbital period for Be star spin-orbit misalignments $\gtrsim 70^\circ$ as a result of weak tidal truncation. 
Most Be/X-ray binaries are redder at optical maximum when the disc is viewed closest to face-on since the disc adds a large red component to the emission. However, A0538-66 is redder at optical minimum. This opposite behaviour requires an edge-on disc at optical minimum and a radially narrow disc such that it does not add a large red signature when viewed face-on. For  A0538-66, the misalignment of the disc to the binary orbit must be about $70-80^\circ$ and  the inclination of the binary orbit to the line of sight must be similarly high, although restricted to  $<75^\circ$ by the absence of X-ray eclipses.
\end{abstract}




\section{Introduction}

Binaries composed of a rapidly  rotating Be type star \citep{Slettebak1982,Porter1996,Porter2003,Rivinius2013} with a neutron star companion are known as Be/X-ray binaries \citep{Negueruela1998,Coe2005,liu2005,haberl2016}. The Be star expels a viscous Keplerian decretion disc from its equator \citep{Pringle1991,Lee1991,Hanuschik1996,Rivinius2006,Carciofi2011}. The Be star spin is often misaligned with respect to the binary orbit  \citep{Hughes1999,Hirata2007,Coe2008,Martinetal2011} as a result of the supernova kick that formed the neutron star \citep{Brandt1995,Martinetal2009b,Salvesen2020}.

Superorbital periods have long been observed in the brightness of Be/X-ray binaries \citep[e.g.][]{Ogilvie2001b,McGowan2008,Rajoelimanana2011}. They are a result of the cooler gas in the disc passing in front of our line of sight to the Be star. The brightness variations therefore depend upon the viewing angle of the system.  However,  the driving mechanism for the periodic behaviour has remained somewhat elusive. The best studied binary, A0538-66 \citep{White1978,Johnston1979,Johnston1980,Skinner1980,Skinner1982,Charles1983,Ducci2016,Ducci2019,Ducci2019b,Ducci2022}, has an orbital period of $P_{\rm orb}=16.6\,\rm day$ and a superorbital period of $t_{\rm super}=421\,\rm day$ \citep{Alcock2001,Rajoelimanana2017} during which the brightness varies by about $0.6\, \rm mag$ \citep[e.g.][]{Alcock1996,Alcock1999,Rivinius2013}.

 Recently, \cite{Martin2023super} showed that superorbital periods may be driven by a nodally precessing disc that is locked to the equator of the spin axis of a Be star, at least for short orbital period binaries. The spin axis of the Be star precesses about the binary angular momentum vector \citep[e.g.][]{Lai1993,Lai1994,Lai2014,Zanazzi2018}. For the Be/X-ray binary A0538–66, the precession of the Be star spin axis occurs on a timescale similar to twice the observed superorbital period \citep{Martin2023super}. The brightness then varies over the superorbital period as the cool disc moves in front of the star \citep[e.g.][]{Alcock2001,Rajoelimanana2011}. The minimum brightness  occurs when the disc is seen close to edge-on (that occurs twice per precession timescale) and covering a large fraction of the star. The magnitude of the superorbital variations therefore depends upon the viewing angle of the system.

X-ray outbursts occur when material flows from the Be star disc and is accreted by the neutron star \citep[e.g.][]{Negueruela2001,Okazaki2001,Okazaki2002}. The X-ray outbursts that occur on the orbital period at periastron passage of the binary are also dependent upon the superorbital phase \citep{Alcock2001}. 
 In A0538-66, the X-ray outbursts are brightest during the dim part of the superobrital period 
\citep{McGowan2003}. When the star is at maximum brightness, the outbursts do not occur or are significantly weakened \citep{Rajoelimanana2017}. The outbursts occur during the optical minimum.  This led to the suggestion that the disc forms and then is depleted on the superorbital period \citep{McGowan2003}. However, an underlying theoretical mechanism to turn the outflow on and off periodically has not been suggested.

In addition to the brightness variation, it has long been recognised that Be stars undergo cyclic variation of their  colour \citep[e.g.][]{Rajoelimanana2011}.  Observations of A0538-66 show that the source is bluer at optical maximum and redder at optical minimum with the colour reddening from V-R of about $-0.1$ to about $0$ during the increase in brightness \citep{Alcock2001}.
However, this is the opposite behaviour to most other Be/X-ray binaries that are redder during optical maximum \citep{Rajoelimanana2011}.

While the precessing disc model can explain the observed superorbital periods in the brightness, it has not been shown how this fits in with the observations of the X-ray outbursts or colour variations. In this Letter, we show that, in the model, the magnitude of the accretion outbursts on to the neutron star  varies over the superorbital period for a sufficiently large misalignment of the Be star spin axis to the binary orbit.    In Section~\ref{simulation} we describe hydrodynamical simulation results for varying misalignment and show how the accretion rate on to the neutron star varies. In Section~\ref{v/r} we discuss how the precessing disc model can explain observations of the colour variations and use this to constrain the misalignment of the Be star disc to the binary orbit, and the inclination of the binary to the line of sight. We draw our conclusions in Section~\ref{concs}.

\section{Variation of the accretion rate on to the neutron star }
\label{simulation}

In this section we first describe the set up of the hydrodynamical simulations. We then consider how the accretion rate on to the neutron star varies over the superorbital period for varying Be star spin axis misalignment. Finally we discuss the cause of the variation.

\begin{table}
\centering
\begin{tabular}{ccccc}
  \hline
{Name}  & $i/^\circ$ &  $\dot M_1/\dot M_{\rm inj}$ & $\dot M_2/\dot M_{\rm inj}$ & $\left<\dot M_2\right>_{\rm min}/\left<\dot M_2\right>_{\rm max}$  \\
\hline
\hline
sim10 & 10  & 0.756 & 0.207 & 0.98 \\
sim30 &  30  & 0.825 & 0.123 & 0.81 \\
sim50 &  50  & 0.906 & 0.054 & 0.66 \\
sim70 &  70  & 0.945 & 0.020 & 0.08\\
\hline
\end{tabular}
\caption{The columns show the simulation name, the misalignment of Be star spin axis relative to the binary orbit, the average accretion rate over the full simulation onto the Be star and onto the neutron star. The final column shows the relative variation over the superorbital period of the orbital-period averaged accretion rate. }
\label{tab:runs}
\end{table}

\begin{figure}
    \centering
    \includegraphics[width=\columnwidth]{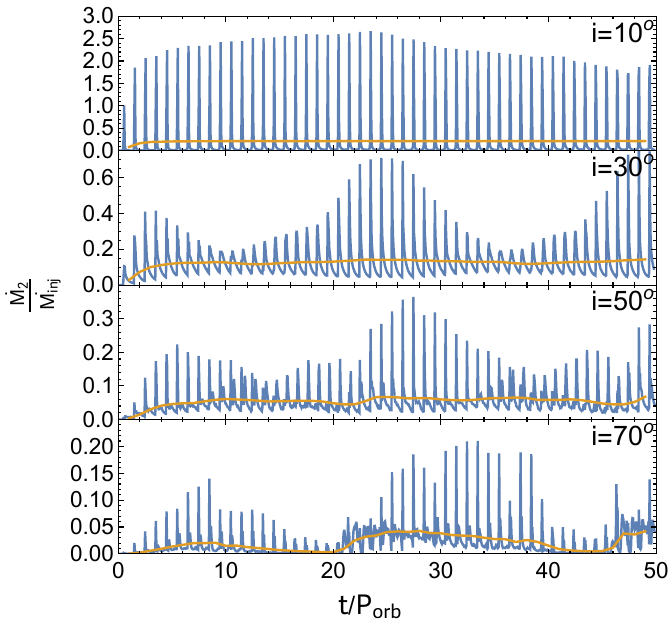}
    \caption{The accretion rate on to the neutron star as a function of time. The misalignment of the spin axis of the Be star relative to the binary orbit is $i=10$, 30, 50, and $70^\circ$ from top to bottom. The blue lines show the accretion rate averaged over intervals of $0.05\,P_{\rm orb}$ while the yellow lines show the accretion rate averaged over the orbital period. Note that the panels have different scales for the accretion rate.  }
    \label{fig:mdot}
\end{figure}

\subsection{Simulation set-up}

To model the Be/X-ray binary we follow the methods of \cite{Martin2023super}. We use the smoothed particle hydrodynamics (SPH) code {\sc phantom} \citep{Price2010,Price2012a,Price2018}. The  Be star has a mass of $M_1=8.84\,\rm M_\odot$  which allows for the canonical neutron star mass of $M_2=1.44\,\rm M_\odot$ \citep{Rajoelimanana2017}.  The binary components are in an orbit with semi-major axis of $a_{\rm b}=59.4\,\rm R_\odot$ and eccentricity $e_{\rm b}=0.72$ and begins at apastron separation. The periastron separation of the binary components is $16.6\,\rm R_\odot$. The binary components are treated as spherical sink particles that accrete the mass and angular momentum of any gas particles that pass inside of their radii \citep{Bateetal1995}. The Be star has an accretion radius of $8\,\rm R_\odot$ and  the neutron star has an accretion radius of  $1\,\rm R_\odot$. Gas particles have a mass of $5\times 10^{-15}\,\rm M_\odot$.  Material is injected in to the disc with Keplerian velocity in a ring at a radius of $R_{\rm inj}=10\,\rm R_\odot$ at a rate of $\dot M_{\rm inj}=10^{-8}\,\rm M_\odot\,yr^{-1}$ \citep[see also e.g.][]{Okazaki2002,Cyr2017,Brown2019,Suffak2022}. The large injection radius increases the resolution of the simulation and prevents implausibly large accretion rates back on to the Be star \citep{Nixon2020,Nixon2021}. The spin axis of the Be star is tilted by misalignment $i$ to the binary angular momentum vector and precesses about the binary angular momentum vector. Following the methods described in \cite{Martin2023super}, the equator of the Be star precesses on a timescale of $t_{\rm prec}=2t_{\rm super}=50.7\,P_{\rm orb}$ in order to fit the observed superorbital period in A0538-66. 

The precession period for the Be star spin-axis formally depends upon the inclination of the star relative to the binary orbit \citep[see equation 2 in][]{Martin2023super}. However, we assume that the superorbital period is fixed between the different inclinations that we consider since we focus on the A0538-66 system. The stellar precession period also depends upon a number of parameters such as the binary masses, the stellar radius, properties of the stellar interior and the spin frequency of the star  \citep{Lai1993,Lai1994} that are not well constrained for A0538-66.  

The \cite{SS1973} viscosity parameter for Be star discs  is typically $\alpha \approx 0.1-0.3$  \citep{Jones2008,Rimulo2018,Ghoreyshi2018,Granada2021}, as is expected for a fully ionised disc \citep{Martin2019}. In the code, the viscosity is implemented by adapting the artificial SPH viscosity \citep{Lodato2010} with $\alpha_{\rm AV}=1$ and $\beta_{\rm AV}=2$ \citep[e.g.][]{Okazaki2002}. The disc is globally isothermal and we take the disc aspect ratio to be $H/R=0.04$ at the stellar radius $R=8\,\rm R_\odot$. In sim30, the density averaged smoothing length in the quasi-steady disc is $\left<h\right>/H\approx 0.5$ which is equivalent to a \cite{SS1973} $\alpha \approx 0.05$. The disc mass and hence resolution increases with the spin-orbit misalignment $i$.

\subsection{Effect of the spin axis misalignment}

\begin{figure*}
    \centering
    \includegraphics[width=2\columnwidth]{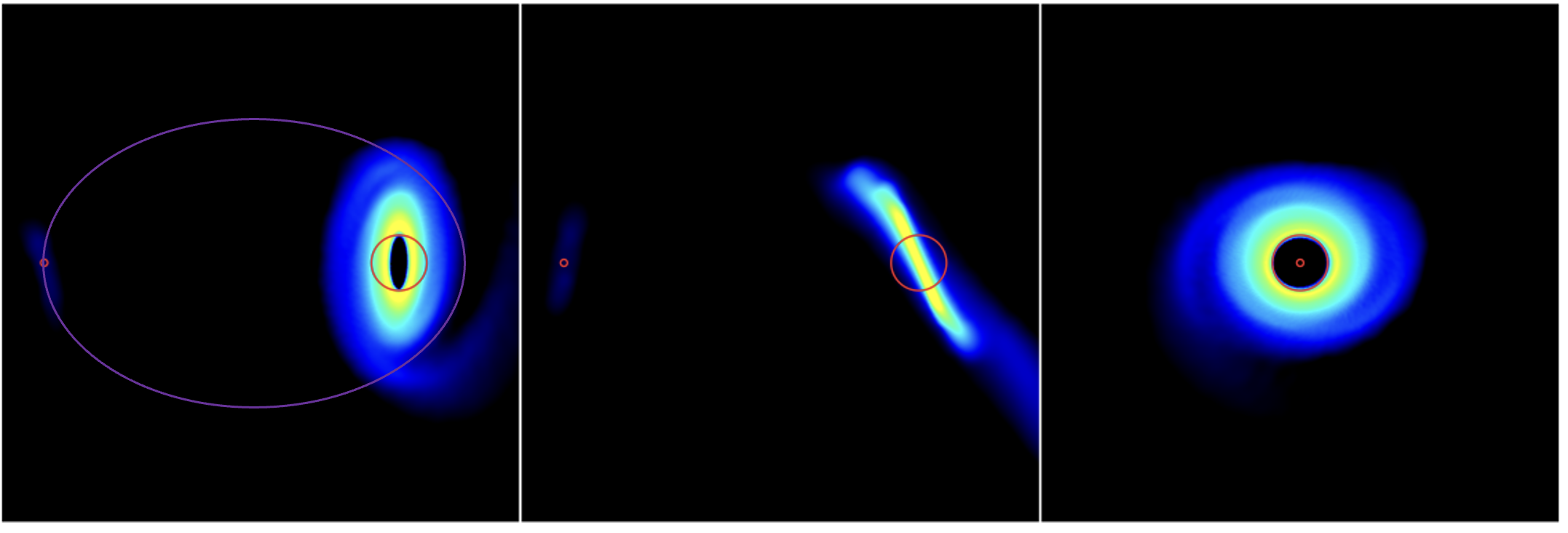}
     \includegraphics[width=2\columnwidth]{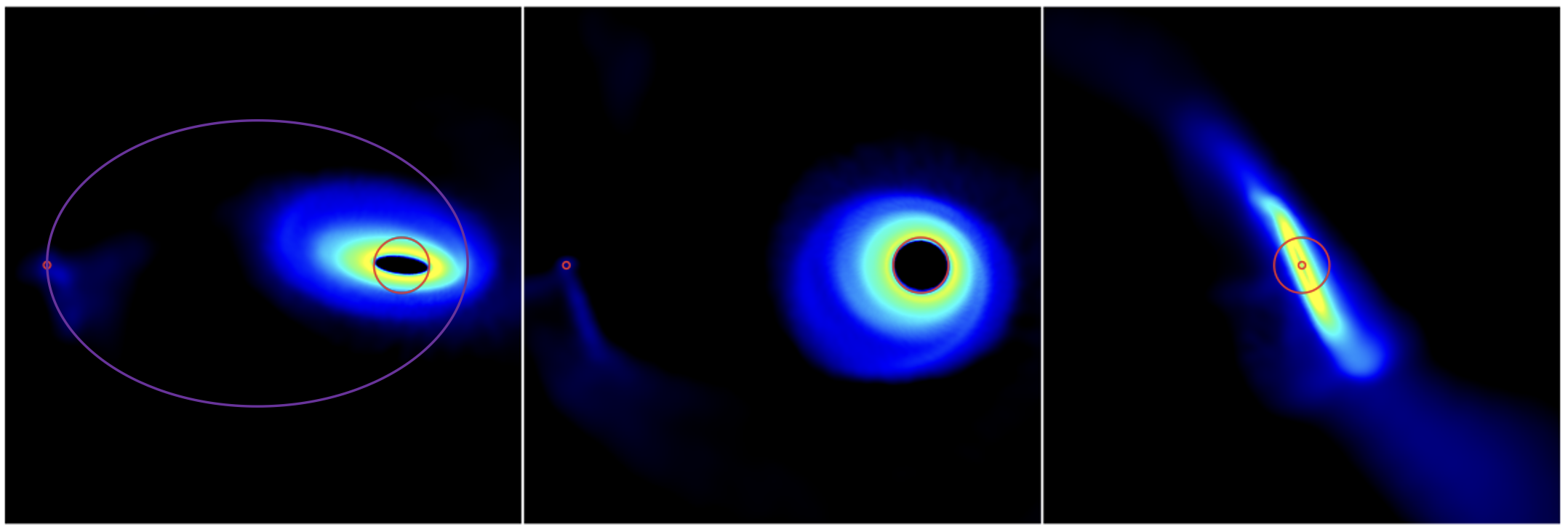}
    \caption{The orientation of the disc viewed in the $x-y$ binary orbital plane (left),  the $x-z$ plane (middle) and the $y-z$ plane (right) for sim70 that has a Be star spin orbit misalignment of $i=70^\circ$. The times shown are $t=13\,P_{\rm orb}$ (upper panels) and $t=25\,P_{\rm orb}$ (lower panels). The binary is at apastron at both times. In each panel, the open red circles show the Be star (right) and the neutron star (left) scaled to their sink radius.  The gas disc is around the Be star. The orbit of the neutron star is shown in magenta in the left panels.The binary orbit is viewed edge-on in the middle and right panels. The colour of the gas shows the column density with yellow being about two orders of magnitude higher than blue. }
    \label{fig:splash}
\end{figure*}

The blue lines in Fig.~\ref{fig:mdot} show the accretion rate on to the neutron star as a function of time for varying misalignment of the Be star spin axis relative to the binary orbit. The yellow lines show the same accretion rate but averaged over the orbital period. The simulations are run for one spin axis precession period, that corresponds to two superorbital periods. Table~\ref{tab:runs} show the parameters for each simulation. Note that sim30 with misalignment $i=30^\circ$ is the same simulation as presented in \cite{Martin2023super}. The third and fourth columns show the average accretion rate on to the Be star and the neutron star, respectively, averaged over the entire simulation. The higher the misalignment, the larger the accretion rate back on to the Be star and the smaller the average accretion rate on to the companion neutron star.

The upper panel of Fig.~\ref{fig:mdot} shows a low misalignment of $i=10^\circ$. Once a quasi-steady state is reached, there is little variation in the orbital-period averaged accretion rate over the superorbital period (yellow line). However, each periastron passage there is a large spike in the accretion rate and therefore an X-ray outburst (see the blue line that is averaged over $0.05\,P_{\rm orb}$). 
For larger misalignment,   there is more variation of the accretion rate over the superorbital period.  For $i=30^\circ$ (second row), the average accretion rate (yellow line) is still quite constant in time, but the peak outburst accretion rate varies (blue line). The longitude of ascending or descending node of the disc is aligned to the binary periapsis at times $t=0$, 25.3 and $50.7\,P_{\rm orb}$. These correspond to times with larger accretion outbursts. The neutron star approaches closest to the disc at these times and is able to accrete more material.

The lower two panels in Fig.~\ref{fig:mdot} show larger misalignments of $i=50^\circ$ and $70^\circ$. Similar behaviour occurs as for $i=30^\circ$ in terms of the peak outburst rate. However,  the variation in the orbital-period averaged accretion rate over the superorbital period increases with misalignment (yellow lines). For $i=70^\circ$, the orbital-period averaged accretion rate varies by about an order of magnitude over the superorbital period and the peak accretion rate has an even larger variation. 
The final column in Table~\ref{tab:runs} shows the ratio of the minimum to maximum orbital period averaged accretion rate over the superorbital period.  There is a clear transition that for $i\gtrsim 70^\circ$, the variation over the superorbital period becomes significant. 

Fig.~\ref{fig:splash} shows the disc with $i=70^\circ$ (sim70) from three different viewpoints at two different times.\footnote{Note that the colour scale is the same as in Fig.~3 in \cite{Martin2023super} and comparing to this figure, the more highly misaligned disc is more radially extended and denser.}  When the disc orientation is such that the longitude of ascending node, or the longitude of the descending node is aligned with the binary periapsis (lower panels), then the accretion rate is at its maximum. When the disc is orientated perpendicular to this (upper panels), the peak accretion rate is at its minimum. The accretion rate during this time has a double peaked appearance since the neutron star approaches the disc twice during each periastron passage \citep[see also][]{Corbet1997,Rajoelimanana2017}.

X-ray outbursts in A0538-66 occur only during the optical minimum part of the superorbital period. In this Section we have shown that X-ray outbursts are largest when the disc longitude of ascending node or descending node is aligned to the binary periapsis (e.g. lower panels of Fig.~\ref{fig:splash}). This suggests that we are viewing the system such that the disc is edge-on at this time.  We discuss further constraints on the binary orientation in Section~\ref{v/r}.

\subsection{Superorbital variations caused by disc truncation}

\begin{figure}
    \centering
    \includegraphics[width=\columnwidth]{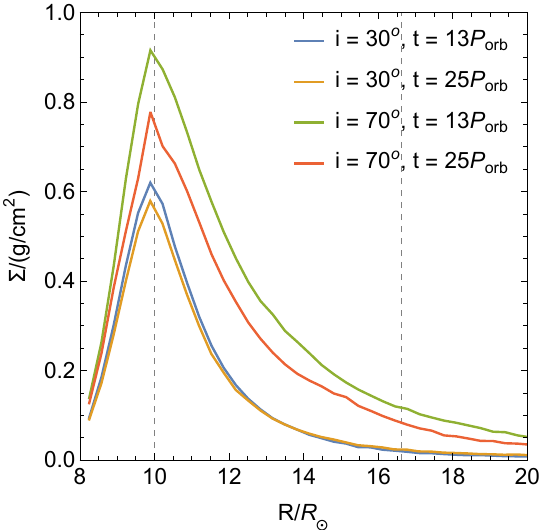}
    \caption{Azimuthally averaged surface density of the disc as a function of radius for two simulations with misalignments $i=30^\circ$ and $i=70^\circ$ each at two different times of $t=13\,P_{\rm orb}$ and $t=25\,P_{\rm orb}$. The left vertical dashed line shows where material is injected into the disc. The right dashed line shows the periastron separation of the binary. Note that the surface density scales with the injection accretion rate.}
    \label{fig:sigma}
\end{figure}

Fig.~\ref{fig:sigma} shows the azimuthally averaged  surface density at times of high peak accretion rate ($t=25\,P_{\rm orb}$) and low peak accretion rate  ($t=13\,P_{\rm orb}$) for sim30 and sim70.  Note that because the disc mass is much less than the binary mass and we do not include disc self-gravity into our calculations, the surface density scales with the input accretion rate, $\dot M_{\rm inj}$.

When tidal truncation is efficient, the size of the disc does not depend upon its longitude of ascending node. 
This can be seen in the surface density profile of the disc simulation with $i=30^\circ$ in Fig.~\ref{fig:sigma}. The surface density profile is very similar at times $t=13\,P_{\rm orb}$ and $t=25\,P_{\rm orb}$. The majority of the mass in the disc is found at radii $\lesssim 14\,\rm R_\odot$. 
Therefore the size of the disc is roughly constant over the precession period and hence the superorbital period. This is why for a misalignment of $i=30^\circ$, the orbital-period averaged accretion rate is constant in time. 

On the other hand, for higher disc misalignment, tidal truncation becomes weaker and the disc spreads out to larger radii \citep{Lubowetal2015,Miranda2015}. Instead, the disc is truncated through close interaction with the neutron star \citep[see also][]{Martin2023cpd}. This is seen in Fig.~\ref{fig:sigma} since the disc surface density is always higher for the more highly inclined disc. Furthermore, there is a significant amount of material flowing out past the binary periastron separation for this disc. The surface density and total mass in the disc is dependent on the disc longitude of ascending node (compare the red and green lines in Fig.~\ref{fig:sigma}).  The neutron star passes closest to the disc when the longitude of ascending node or descending node is aligned to the binary periapsis ($t=25\,P_{\rm orb}$, and see the lower panels of Fig.~\ref{fig:splash}). The surface density (and disc mass) is therefore lower during this time and the accretion rate on to the neutron star is higher. The variable truncation of the disc over the superorbital period is why the average accretion rate  has a large variation over the superorbital period for the more highly inclined discs. 

 The time of the largest accretion outbursts occurs later for larger misalignments (see Fig.~\ref{fig:mdot}). For $i=30^\circ$, it occurs around $t=25\,P_{\rm orb}$ because the disc is locked to the equator of the Be star and at this time, the longitude of ascending node is aligned to the binary periapsis. However, for larger misalignments, the disc does not remain exactly locked to the Be star equator. The larger and more massive disc begins to dominate the dynamics compared to the accretion torque from the new material being added. In the absence of the accretion of material in to the disc, the disc would precess on a timescale dictated by the binary torque \citep{Lubow2000,Bateetal2000}.

A highly misaligned disc can undergo von-Zeipel--Kozai--Lidov \citep{vonZeipel1910,Kozai1962,Lidov1962} oscillations of eccentricity and inclination \citep{Martinetal2014,Martinetal2014b,Fu2015,Lubow2017,Zanazzi2017,Franchini2019}. While we do see some eccentricity growth in the disc (see Fig.~\ref{fig:splash}), this is not attributed to ZKL oscillations because there is no corresponding inclination change. We do not see evidence for such oscillations in these simulations because the disc is so narrow that the timescale for material to flow through the disc is shorter than the ZKL timescale \citep{Smallwood2021,Smallwood2023}. There is some evidence of disc warping seen in Fig.~\ref{fig:splash} that is possible because of the larger radial extent of the disc for high misalignment.

\section{Variation of the  colour }
\label{v/r}

For a misaligned and nodally precessing disc, large variations in the colour may occur because (i) the stellar emission has a varying amount of obscuration by the disc and (ii) the magnitude of the disc contribution depends upon its size and its inclination to the line of sight.  The colour variations observed depend on how the inclination of the disc to the line of sight varies over the nodal precession. 

\subsection{Variation of the disc inclination to the line of sight}

The inclination of the angular momentum vector of the binary  relative to the line of sight, $i_{\rm b-los}$, is fixed for an isolated system with a disc mass that is small compared to the binary mass. We assume that the misalignment of the disc angular momentum vector to the binary angular momentum vector, $i$, is constant and the disc is not warped.  However, the inclination of the angular momentum vector of the disc relative to the line of sight, $i_{\rm d-los}$, oscillates over the nodal precession period as the disc angular momentum vector precesses about the binary angular momentum vector. The disc is viewed face-on for $i_{\rm d-los}=0^\circ$ and edge-on for $i_{\rm d-los}=90^\circ$, and similarly for the binary. 

\begin{figure}
    \centering
    \includegraphics[width=\columnwidth]{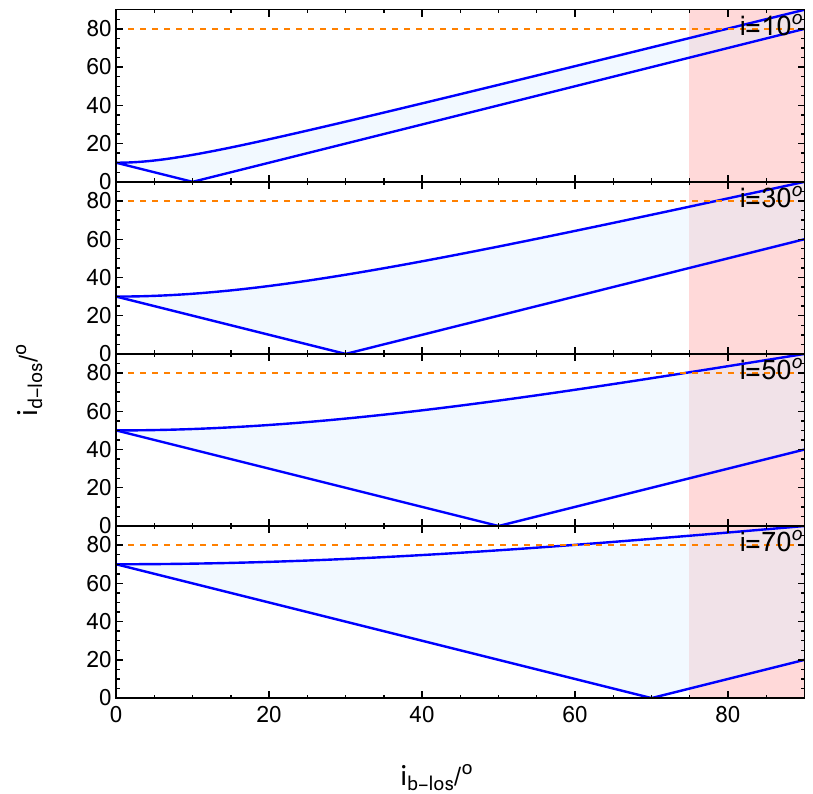}
    \caption{The range of the inclination of the disc angular momentum vector to the line of sight, $i_{\rm d-los}$ as a function of the inclination of the binary angular momentum vector to the line of sight, $i_{\rm b-los}$ for different binary-disc misalignments, $i$. The maximum disc inclination (upper blue lines) occurs at optical minimum when the disc is closest to edge-on. The minimum disc inclination (lower blue lines) occurs when the disc is closest to face-on at optical maximum. The red shaded region shows the prohibited range for $i_{\rm b-los}$ from the lack of X-ray eclipses in A0538-66. The orange dashed lines show the minimum value for $i_{\rm d-los}$ at optical minimum  for a shell profile in A0538-66.  }
    \label{range}
\end{figure}

Fig.~\ref{range} shows the range of the disc inclination relative to the line of sight, $i_{\rm d-los}$, over a nodal precession period at fixed values of the binary inclination relative to the line of sight, $i_{\rm b-los}$. The wider the vertical shaded range, the larger the colour change of the star may be. 
In A0538-66, the absence of X-ray eclipses means that the inclination of the binary to the line of sight must be $i_{\rm b-los}<75^\circ$ \citep{Skinner1980,Rajoelimanana2017}. The unphysical region is shaded in red in Fig.~\ref{range}.

In the precessing disc model, the disc is closest to edge-on ($i_{\rm d-los}=90^\circ$) at optical minimum and closest to face-on ($i_{\rm d-los}=0^\circ$) at optical maximum. 
The presence of the shell profile in the H$\alpha$ line suggests that we are viewing the disc edge-on at optical minimum \citep{Hummel2000}. During the minimum brightness, the inclination of the disc to the line of sight must be large, $i_{\rm d-los}\gtrsim 80^\circ$ \citep[][]{Rajoelimanana2017}. The orange dashed lines  in each panel of Fig.~\ref{range} show $i_{\rm d-los}=80^\circ$. The maximum value for $i_{\rm d-los}$ (upper blue line) must be above this line. This suggests that we require the misalignment of the disc to the binary orbital plane to be $i>50^\circ$.  

There is however, a further constraint. During the bright part of the superorbital period, the star must be seen almost unobscured \citep{Alcock2001}. This means that we require $i_{\rm d-los}$ to be close to zero. Therefore, the lower blue line in Fig.~\ref{range} needs to be close to zero. To satisfy all of these constraints, requires $i\approx i_{\rm b-los}$ since this gives the largest range of values for $i_{\rm d-los}$ over a precession period.  These three  constraints, namely, an edge-on disc at optical minimum, a face-on disc at optical maximum and $i_{\rm b-los}<75^\circ$, are all satisfied for a disc that is misaligned to the binary orbit by about $i\approx 70-80^\circ$. The inclination of the binary to the line of sight must also have a similar value, but $i_{\rm b-los}$ is constrained to be $<75^\circ$ by the lack of X-ray eclipses.

\subsection{Reddening of the star}

 There are two effects that lead to reddening of the star. First, when the disc is seen edge-on, it obscures more of the star making it appear redder \citep{Rajoelimanana2017}.  The magnitude of the {\it reddening at optical minimum} depends upon the inclination of the disc to the line of sight at optical minimum.  Secondly, when the disc is seen closer to face-on, its (red) light adds to the light from the star and makes it appear redder.  This {\it reddening at optical maximum} depends upon both the inclination of the disc to the line of sight and the radial extent of the disc. The overall variation during the disc precession depends upon the relative magnitude of these two effects. Both of these effects, namely the amount of emission from the disc and the amount of obscuration of the Be star, are very dependent on the disc density and the wavelength regime.   The disc does not add or obscure much in the UV but greatly contributes to the  infrared, as well as some contribution in the visible. Note that the density of the disc in our simulations scales with the mass injection rate, which is unknown.

A disc with a very narrow radial extent (because of a small periastron separation, such as A0538-66) may not add a significant amount of red light even when it is face-on. However, the star can appear redder at optical minimum if the disc is highly inclined to the line of sight since stellar obscuration dominates the reddening. This type of behaviour requires a large binary-disc misalignment, $i$, a large inclination of the binary to the line of sight, $i_{\rm b-los}$ and a radially narrow disc. A disc with a larger periastron separation has a larger radial extent and may appear redder at optical maximum. 

The outcome for a particular system depends upon the misalignment of the disc relative to the binary orbit, the viewing angle and the radial extent of the disc.
Detailed calculations of the effects on the star's colour variations are beyond the scope of this work, but should be investigated in future \citep[see, for example][]{Suffak2023}. Note that in this Section we have  assumed that the disc is not warped over its radial extent. This is reasonable for systems with a small orbital period but may be important for longer period systems \citep[e.g.][]{Martinetal2011}. Disc warping may lead to more obscuration than predicted by a flat but tilted disc model.

\section{Conclusions}
\label{concs}

We have examined the precessing disc model for driving superorbital periods in Be/X-ray binaries. Material is ejected from the equator of the Be star and forms a decretion disc. The spin axis of the star is misaligned relative to the binary orbit and undergoing spin axis precession. For short orbital period binaries, the disc precesses on the timescale of the spin axis precession of the Be star. Aside from the brightness variations that were explained in \cite{Martin2023super} we have shown how the model can explain two additional features that change on the superorbital period, namely the observed X-ray outbursts and the observed  colour changes.

 We have demonstrated that the accretion rate on to the neutron star  may change by an order of magnitude over a precession timescale, and therefore over a superorbital period. Since X-ray outbursts in A0538-66 occur only during the optical minimum part of the superorbital period, this suggests that the orientation of the system is such that we see the disc edge-on when its longitude of ascending or descending node is close to the binary periapsis.   The disc must be misaligned by $i\gtrsim 70^\circ$ to the binary orbital plane in order for large variation in the average accretion rate to occur. Tidal truncation is inefficient for highly misaligned discs and the truncation of the disc depends upon the disc nodal alignment with the binary periapsis.

The precessing disc model naturally explains the colour variations of Be/X-ray binaries over the superorbital period. Systems that have a high disc inclination to the line of sight (meaning that the disc is close to edge-on) at optical minimum and a small periastron separation (like  A0538-66) become redder at optical minimum when the disc passes in front of the star. At times when the disc is closer to face-on, the narrow radial extent of the disc does not add a large red signature to the emission. However, systems with a larger periastron separation can have a radially wider disc that emits a significant amount of red light meaning that the star is redder at maximum brightness.  
Observations of the brightness and colour of A0538-66  can be explained by a disc that is misaligned to the binary orbital plane by  $i\approx 70-80^\circ$. The inclination of the binary to the line of sight must have a similarly high value but is constrained to $<75^\circ$ by the lack of X-ray eclipses.

\section*{Acknowledgements}
We are grateful to an anonymous referee for useful comments. Computer support was provided by UNLV’s National Supercomputing Center. RGM acknowledges support from NASA through grant 80NSSC21K0395. We acknowledge the use of SPLASH \citep{Price2007} for the rendering of Fig.~\ref{fig:splash}.

\section*{Data Availability}

 The data underlying this article will be shared on reasonable request to the corresponding author.
 



\bibliographystyle{mnras}
\bibliography{mainmnras} 








\bsp	
\label{lastpage}
\end{document}